\documentclass[aps,pra,reprint,twocolumn,superscriptaddress,amsmath,amssymb,amsfonts,floats,floatfix,noraggedbottom,nobalancelastpage,dvips,showpacs,10pt]{revtex4-1}

\usepackage{epsfig} 
\usepackage{psfrag}
\usepackage{braket}
\usepackage{color}
\usepackage{textcomp}
\usepackage{graphicx}
\usepackage{latexsym}
\usepackage{hyperref}
\hypersetup{dvips,
pdfauthor={Juan Carrasquilla, Andrea Di Ciolo, Federico Becca, Victor Galitski, and Marcos Rigol},
  pdftitle={Nature of the phases in the frustrated XY model on the honeycomb lattice},
  colorlinks=true,
  linkcolor=blue,
  citecolor=blue,
  pdfpagemode=UseNone
}

\usepackage{breakurl}

\newcommand{\bbraket}[1]{\braket{\hspace{-2pt}\braket{#1}\hspace{-2pt}}}

\begin{document}

\title{Nature of the phases in the frustrated $XY$ model on the honeycomb lattice}

\author{Juan Carrasquilla} 
\affiliation{Department of Physics, The Pennsylvania State University,
University Park, Pennsylvania 16802, USA}
\affiliation{Department of Physics, Georgetown University, Washington, DC 20057, USA}
\author{Andrea Di Ciolo} 
\affiliation{Joint Quantum Institute and Department of Physics, University of Maryland, 
College Park, Maryland 20742, USA}
\affiliation{Department of Physics, Georgetown University, Washington, DC 20057, USA}
\author{Federico Becca}
\affiliation{Democritos Simulation Center CNR-IOM Istituto Officina dei Materiali, Trieste, Italy} 
\affiliation{International School for Advanced Studies (SISSA), Via Bonomea 265, 34136 Trieste, Italy}
\author{Victor Galitski}
\affiliation{Joint Quantum Institute and Department of Physics, University of Maryland, 
College Park, Maryland 20742, USA} 
\author{Marcos Rigol} 
\affiliation{Department of Physics, The Pennsylvania State University,
University Park, Pennsylvania 16802, USA}

\begin{abstract}
We study the phase diagram of the frustrated $XY$ model on the honeycomb lattice 
by using accurate correlated wave functions and variational Monte Carlo simulations. 
Our results suggest that a spin-liquid state is energetically favorable in the region 
of intermediate frustration, intervening between two magnetically ordered phases. 
The latter ones are represented by classically ordered states supplemented with a long-range 
Jastrow factor, which includes relevant correlations and dramatically improves the 
description provided by the purely classical solution of the model. The construction 
of the spin-liquid state is based on a decomposition of the underlying bosonic particles 
in terms of spin-$1/2$ fermions (partons), with a Gutzwiller projection enforcing 
no single occupancy, as well as a long-range Jastrow factor.
\end{abstract}

\pacs{
  75.10.Kt, 
  67.85.Jk, 
  21.60.Fw, 
  75.10.Jm  
}


\maketitle

A quantum spin liquid is an exotic state in which strong quantum 
fluctuations (usually generated by frustration) preclude ordering 
or freezing, even at zero temperature~\cite{Balents2010}. Despite 
intensive theoretical and experimental research, finding quantum 
spin liquids in materials and in realistic spin models continues 
to be a challenge. A remarkable example where the existence of 
such a state has been inferred is the spin-$1/2$ kagome-lattice 
Heisenberg antiferromagnet, which has been extensively studied 
both theoretically and 
experimentally~\cite{Balents2010,Yan03062011,Han2012}, even though 
the precise nature of the spin-liquid state (gapped vs gapless) is 
still under debate~\cite{Han2012,Messio2012,Depenbrock2012,Iqbal2013}. 
Another model that has recently received considerable attention for 
its potential to realize spin-liquid states is 
the spin-$1/2$ Heisenberg model on the honeycomb lattice, with 
nearest-neighbor (NN) $J_1$ and next-to-nearest neighbor (NNN) 
$J_2$ exchange interactions~\cite{Wang2010,Lu2011,Clark2011,Mosadeq2011,
Albuquerque2011,Bishop2012,Mezzacapo2012,Ganesh2013,Zhu2013,Gong2013}. This 
is in part motivated by its close relation to the Hubbard model, 
for which the possibility of having a spin-liquid ground state has 
been under close scrutiny~\cite{Meng2010,Sorella2012,Assad2013}.

A closely related spin model with a rich phase diagram and the promise
to support a gapless spin liquid phase is the $J_1-J_2$ spin-$1/2$ $XY$ 
model on the honeycomb lattice~\cite{Varney2011,Sedrakyan2013}, which 
is the main subject of this Rapid Communication. Its Hamiltonian can be written as
\begin{equation}
\mathcal{H} = J_1 \sum_{\braket{ij}} ( S_i^x S_j^x + S_i^y S_j^y  ) + J_2
  \sum_{\bbraket{ij}} ( S_i^x S_j^x + S_i^y S_j^y ),
  \label{eq:Ham_xy}
\end{equation}
where $S_i^\alpha$ is the $\alpha$th component of the spin-$1/2$ operator 
at site $i$. This model can be thought of as a Haldane-Bose-Hubbard 
model~\cite{Varney2010,Wang2011,Varney2011,Varney2012}, i.e., the Haldane 
model~\cite{Haldane1988} on the honeycomb lattice with NN hopping 
$J_1$ and complex NNN hopping $|J_2|e^{\imath\phi}$, where spinless fermions 
are replaced by hard-core bosons and $\phi=0$. Hard-core boson creation
and annihilation operators can then be mapped onto spin operators 
($b_i^{\dag}\rightarrow S^{+}_i$, $b_i^{}\rightarrow S^{-}_i$) leading
to Eq.~\eqref{eq:Ham_xy}. The total number of bosons ($N$) is related 
to the total magnetization in the spin language, since $n_i=S^{z}_i+1/2$.
Here, we focus on the half-filled case, where $N$ equals one half the 
number of sites ($V$).

\begin{figure}[!b]
\includegraphics[width=0.4\textwidth,angle=-90]{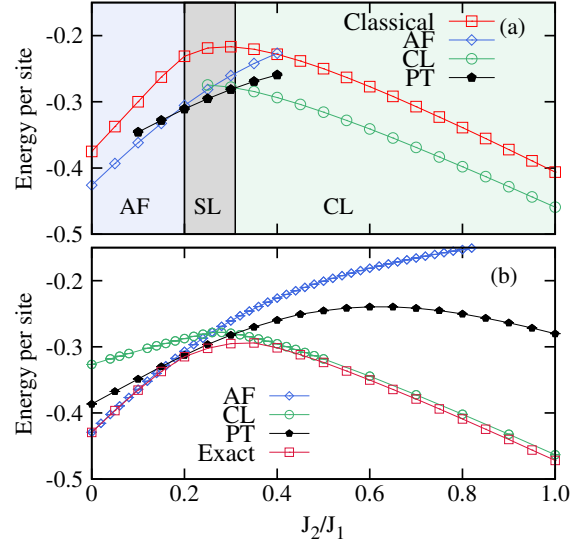}
\caption{(Color online) (a) Extrapolated best energies in the 
thermodynamic limit, as well as the final phase diagram (colored regions 
labeled AF, SL, and CL) based on the properties of the states considered. 
The energy of the purely classical solution of the model is shown for comparison.
(b) Energies of antiferromagnetic and spin-liquid states compared 
to the exact results on the $4 \times 4 \times 2$ cluster.}
\label{fig:phased}
\end{figure}

This model was studied in Ref.~\cite{Varney2011} by means of exact 
diagonalization on small clusters. There, evidence was found 
supporting the existence of a spin liquid surrounded by two magnetically 
ordered states, namely, an antiferromagnetic (collinear) state at 
lower (higher) $J_2/J_1$. The spin-liquid phase was suggested to be gapless 
and characterized by a distinctive parameter dependent feature in the momentum 
distribution $n({\bf k})$, similar to a Bose surface, thus suggesting
the presence of an exotic Bose metal~\cite{Paramekanti2002,
Phillips2003,Motrunich2007,Sheng2009,Mishmash2011,Varney2011,Dang2011,Varney2012}.

Here, we study the phase diagram of the $J_1-J_2$ spin-$1/2$ $XY$
model on the honeycomb lattice using variational Monte Carlo (VMC). By 
utilizing an accurate, yet simple and intuitive representation of magnetically 
ordered states, we find that the system remains in the antiferromagnetic 
(AF) phase for $0\leq J_2/J_1\leq 0.2$, while a collinear 
(CL) state stabilizes for $0.3\lessapprox J_2/J_1 \lessapprox 1.1$. 
The accuracy of our representation, tested against exact diagonalization results 
on small clusters, is unprecedented. In the intermediate region 
$0.2\lessapprox J_2/J_1 \lessapprox 0.3$, we find that the ordered phases 
have a higher energy than a fractionalized partonic state, which is consistent 
with a gapped spin liquid. The variational energy of such a state is 
gauged against a wide range of carefully optimized magnetically ordered 
spin states, as well as states that allow the breaking of spatial symmetries. 
Extrapolations of energy to the thermodynamic limit of 
the best trial states, as well as the phase diagram of the model, are 
presented in Fig.~\ref{fig:phased}. The energy of the purely classical solution 
of the model is shown to make apparent the importance of introducing quantum 
fluctuations in our trial states. 

As trial wave functions for the magnetically ordered states, we use classical 
spin-waves on the $XY$ plane supplemented with long-range Jastrow factors:
\begin{equation}
|\Psi_{\mathbf{Q}}\rangle=\mathcal{J}_z \prod_{i} \left(|\downarrow\rangle_{i} + 
e^{\imath\mathbf{Q}.\mathbf{R}_i+\imath \eta_{\mathbf{R}_i}}|\uparrow\rangle_{i} \right),
\label{eq:spirals}
\end{equation}
where $i$ runs over the positions of the spins, $\mathbf{Q}$ is the wave vector 
of the classical spin wave, and $\eta_{\mathbf{R}_i}$ is the phase shift between 
the two spins within the unit cell. Without loss of generality, we assume 
$\eta_{\mathbf{R}_i}=0$ if $\mathbf{R}_i$ belongs to the sublattice $\mathcal{A}$ 
and an arbitrarily chosen $\eta_{\mathbf{R}_i}=\eta$ if it belongs to the 
sublattice $\mathcal{B}$. The long-range Jastrow factor $\mathcal{J}_z=
\exp\left(\frac{1}{2}\sum\limits_{i,j}v_{ij}S^{z}_i S^{z}_{j}\right)$, with 
$v_{ij}$ to be optimized, is also considered to include relevant (i.e., out-of-plane)
quantum correlations. In the bosonic language, the magnetically ordered states are 
nothing but condensates where particles macroscopically populate 
finite-$\mathbf{Q}$ momentum states. 

The trial wave function for the intermediate spin-liquid state is written as
\begin{equation}
|\Psi_\textrm{SL}\rangle  
 = \mathcal{J}_z P_{G} |\Psi_{c_{\uparrow},c_{\downarrow}}\rangle,
\end{equation}
where $|\Psi_{c_{\uparrow},c_{\downarrow}}\rangle$ is the ground state of 
a free-fermion Hamiltonian
\begin{equation}
\mathcal{H}_{c_{\uparrow},c_{\downarrow}}=\sum\limits_{\langle i,j \rangle} \Psi^{\dag}_i 
T^{ij}_{\scriptscriptstyle  NN} \Psi_j + \sum\limits_{\langle \langle i,j \rangle \rangle} 
\Psi^{\dag}_i T^{ij}_{\scriptscriptstyle NNN} \Psi_j 
+ \sum\limits_{ i} \Psi^{\dag}_i M^{i} \Psi_i,
\label{eq:meanfield}
\end{equation}
in which $\Psi^{\dag}_i=(c^{\dag}_{\uparrow,i},c^{\dag}_{\downarrow,i})$ acts 
at site $i$ and is composed by two fermions (partons) $c^{\dag}_{\uparrow,i}$ and 
$c^{\dag}_{\downarrow,i}$. The latter ones are related to the underlying physical 
hard-core bosons (or spin-$1/2$ operators) via 
$b^{\dag}_i= c^{\dag}_{\uparrow,i}c^{\dag}_{\downarrow,i}$. The matrices
\begin{equation}
T^{ij}_{n}=\left[\begin{matrix}
t^{\uparrow}_{ij} & t^{s}_{ij}+ t^{a}_{ij} \\
t^{s*}_{ij}- t^{a*}_{ij} & t^{\downarrow}_{ij}
\end{matrix}\right],
\,\,\,\,\,M^{i}=\left[\begin{matrix}
m^{\uparrow \uparrow}_i & m^{\uparrow \downarrow}_i \\
m^{\uparrow \downarrow*}_i & m^{\downarrow \downarrow}_i
\end{matrix}\right]
\end{equation}
contain hopping parameters and on-site couplings ($n=NN$ and $NNN$). 
In order to have a state that lives in the correct Hilbert space (with one 
spin per site), a Gutzwiller projector $P_{G}$ must be introduced, imposing 
no single occupancy for fermions. Furthermore, a long-range Jastrow factor 
$\mathcal{J}_z$ is also considered. The Jastrow factors, as well as all 
parameters in Eq.~\eqref{eq:meanfield}, are carefully optimized using VMC 
methods as described in Refs.~\cite{Sorella2005} and~\cite{Yunoki2006}. 
After imposing particle-hole (PH) symmetry in the variational ansatz 
\eqref{eq:meanfield}, the most energetically favorable mean-field state is 
found to have real NN hopping $t^{\uparrow}_{ij}=-t^{\downarrow}_{ij}=t=1$, 
as well as nonvanishing complex on-site $m^{\uparrow \downarrow}_i$ and 
real NNN ``spin-orbit'' hopping $t^{s}_{ij}=t^{s*}_{ij}$, while 
$t^{a}_{i,j}=m^{\downarrow \downarrow}_i=m^{\uparrow \uparrow}_i=0$.
Note that, at half filling, PH symmetry $b_j^{\dag}\rightarrow b_j$ in the 
bosonic language (i.e., $c^{\dag}_{\uparrow,j}\rightarrow-\imath c_{\downarrow,j}$ 
and $c^{\dag}_{\downarrow,j} \rightarrow \imath c_{\uparrow,j}$) implies that both 
$T^{ij}_{n}$ and $M^{i}$ must be written in terms of Pauli matrices, which 
reduces the total number of independent parameters. 

\begin{figure}[!t]
\includegraphics[width=0.34\textwidth,angle=-90]{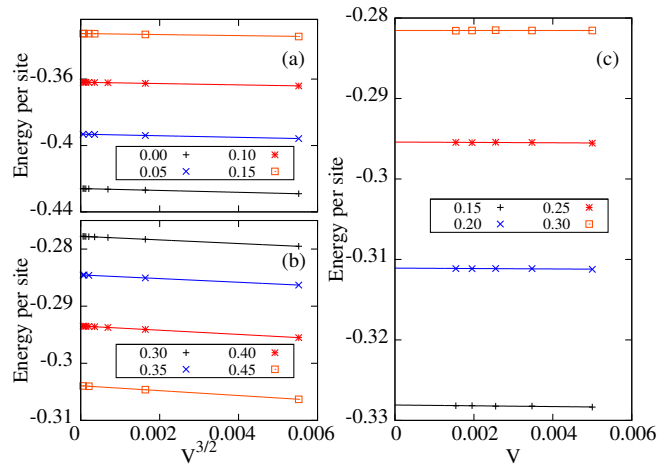}
\caption{(Color online) Finite-size scaling of the AF energies (a), the CL 
states (b), and projected fermionic states (c). In the legends, the numbers 
indicate values of $J_2/J_1$.}
\label{fig:comparison}
\end{figure}

In Fig.~\ref{fig:phased} we present results for the energy vs 
$J_2/J_1$ on a $4\times4\times2$ cluster and compare them with those 
obtained from exact diagonalization. For $J_2/J_1\lessapprox 0.2$ 
($J_2/J_1\gtrapprox 0.35$), the best variational states are given by 
Eq.~\eqref{eq:spirals} with $\mathbf{Q}=\Gamma$ ($\mathbf{Q}=M$) and a 
phase shift $\eta=\pi$ ($\eta=0$). They correspond to antiferromagnetic 
states with collinear order. These two ordered states surround an intermediate 
region where a state constructed from Eq.~\eqref{eq:meanfield} possesses the 
best variational energy. The discrepancy in the energy of our trial states 
with respect to results from exact diagonalization is always less than 
$3\%$, except for $J_2/J_1=0.3$ for which it is $\sim 4\%$ \cite{suppmat}.

For larger cluster sizes, we have analyzed an extensive set of possible 
states. For the classically ordered states, we minimized the energy
for a large number of spin waves generated by nonequivalent $\mathbf{Q}$ 
vectors and for a dense grid of values of $\eta$. For spin-wave states 
with generic pitch vectors $\mathbf{Q}$, it is important to allow for 
Jastrow factors that break rotational symmetry. Regarding the
projected fermionic ansatz, we have considered translationally and 
rotationally invariant states and also allowed for breaking spatial 
symmetries. We have done that by considering enlarged unit cells that 
contain $4$ and $18$ sites, as well as states in which the couplings 
of the mean-field Hamiltonian form plaquette-like structures, and 
fermionic states supplemented with a $\mathcal{J}_z$ such that the
breaking of rotational invariance is allowed. Within that set of 
fermionic trial wave functions, the state with lowest energy, as the 
system size is increased, is such that no spatial symmetries in both 
$\mathcal{J}_z$ and $\mathcal{H}_{c_{\uparrow},c_{\downarrow}}$ 
are broken~\cite{suppmat}.

Within the magnetic states described by the ansatz of Eq.~\eqref{eq:spirals}, 
we find that quantum fluctuations (accounted for by $\mathcal{J}_z$) 
stabilize the antiferromagnetic (AF) phase for larger values of 
$J_2/J_1$ with respect to the classical solution. Indeed, the
$\mathbf{Q}=\Gamma$ ($\eta=\pi$) state remains lower than other spin-wave 
states up to $J_2/J_1 \approx 0.25$ (to be compared to $J_2/J_1=1/6$ for 
the classical solution). Similar results were obtained for the $J_1-J_2$ 
spin-$1/2$ Heisenberg model on the honeycomb lattice~\cite{Ganesh2013}. 
Furthermore, for $0.25 \lessapprox J_2/J_1 \lessapprox 1.1$, the best
magnetic state has collinear (CL) order, with $\mathbf{Q}=M$ and $\eta=0$. 
This outcome is in contrast with the classical limit, where states with 
incommensurate order are found. Most importantly, among all states of the 
form~\eqref{eq:spirals} in the clusters considered, no single quantum 
spin wave has lower energy in the intermediate $J_2/J_1$ region than 
that of the state based on Eq.~\eqref{eq:meanfield}.

The trends discussed above are confirmed by simulations on clusters 
with sizes up to $18\times18\times2$, and in extrapolations of the 
energies to the thermodynamic limit, as presented in Fig.~\ref{fig:phased}. 
We expect finite size corrections to the energy $e_0$ in the AF phase to 
be of the form $e_0\left(V\right)=e_0\left(\infty\right)-c_0/V^{3/2}$
\cite{Fisher1989} [see Fig.~\ref{fig:comparison}(a)] where the slope 
$c_0$ is proportional to the velocity of the AF spin wave. The latter 
(not shown) decreases approximately linearly with increasing 
frustration (for small frustration), as in the AF phase of the 
frustrated Heisenberg model on the same geometry~\cite{Mattsson1994}. 
The aforementioned scaling relation describes the data in the CL phase 
as displayed in Fig.~\ref{fig:comparison}(b). Finally, for the projected 
fermionic state, we assume the simple form 
$e_0\left(V\right)=e_0\left(\infty\right)-d_0/V$ to extract 
$e_0\left(\infty\right)$ as reported in Fig.~\ref{fig:comparison}(c).

\begin{figure}[t]
\includegraphics[width=0.23\textwidth,angle=-90]{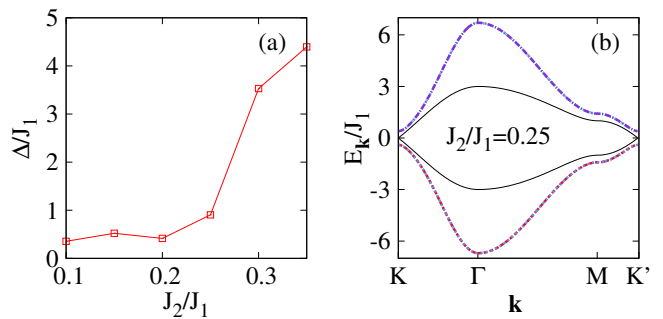}
\caption{(Color online) (a) Single-particle gap of the optimized fermionic 
Hamiltonian in Eq.~\eqref{eq:meanfield} as a function of $J_2/J_1$ for the 
$18\times18\times2$ cluster. (b) The unprojected band structure for 
$J_2/J_1=0.25$. For comparison, black solid lines show the band structure 
for real NN hopping only.}
\label{fig:gap_bands}
\end{figure}

In what follows, we study the properties of the 
projected fermionic state that bears the lowest energy at intermediate 
frustration. After minimizing the energy on several clusters, 
the resulting ansatz $|\Psi_{c_{\uparrow},c_{\downarrow}}\rangle$ 
was found to have a gap to single-particle fermionic excitations 
$\Delta$ as shown in Fig.~\ref{fig:gap_bands}(a) for a $18\times18\times2$ 
cluster. The corresponding unprojected band structure for $J_2/J_1=0.25$ 
(four bands) is shown in Fig.~\ref{fig:gap_bands}(b).

\begin{figure}[!t]
\includegraphics[width=0.41\textwidth,angle=-90]{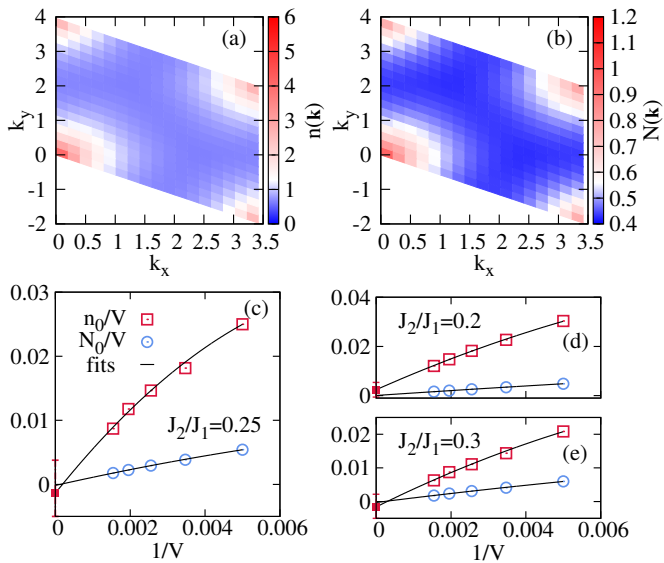}
\caption{(Color online) Momentum distribution $n\left(\mathbf{k}\right)$ (a) 
and Static structure factor $N\left(\mathbf{k}\right)$ (b) for $J_2/J_1=0.25$ 
on the $18\times18\times2$ cluster. Finite size scaling of the condensate 
fraction $n_0/V$ and $N_0/V$ for $J_2/J_1=0.25$ (c), $J_2/J_1=0.2$ (d), 
and $J_2/J_1=0.3$ (e). The solid lines correspond to fits of the data to a 
second-order polynomial. The corresponding extrapolation of the condensate fraction 
is shown with filled red squares.}
\label{fig:observables}
\end{figure}

Whether the system possesses magnetic order is assessed calculating 
the momentum distribution function $n^{\alpha,\beta}\left(\mathbf{k}\right)$, 
where $\alpha$ and $\beta$ denote the two sites inside the unit cell 
[i.e., $n^{\alpha,\beta}\left(\mathbf{k}\right)$ is a $2\times2$ matrix]. 
In Fig.~\ref{fig:observables}(a), we report the trace 
$n\left(\mathbf{k}\right)=\operatorname{tr}[n^{\alpha,\beta}\left(\mathbf{k}\right)]$ 
of this matrix for $J_2/J_1=0.25$. We find a clear peak at the $\Gamma$ 
point, which might suggest that the resulting state is still antiferromagnetically 
ordered. However, upon a finite-size scaling analysis of the condensate fraction
$n_0/V$, where $n_0$ is the largest eigenvalue of the one-body density matrix 
($\langle b^{\dag}_i b^{}_j \rangle$ in the bosonic language or
$\langle S^{+}_i S^{-}_j \rangle$ in the spin language), we find that the state 
is not ordered in the thermodynamic limit. Figure~\ref{fig:observables}(c) shows 
the evolution of $n_0/V$ for $J_2/J_1=0.25$, as the system size is increased. 
The solid black line corresponds to a fit to a second-order polynomial,
which makes apparent that $n_0/V$ vanishes as $V\rightarrow\infty$ (within the error 
bar of the fit). Figures \ref{fig:observables}(d) and \ref{fig:observables}(e)
also make apparent that the same happens in the boundaries of the region where
the partonic state has the lowest energy. 

We have also evaluated the structure factor $N^{\alpha,\beta}\left(\mathbf{k}\right)$
(in which the constant $\mathbf{k}=0$ term has been subtracted), to look for diagonal 
ordering in our projected fermionic ansatz. Results for the trace of 
$N\left(\mathbf{k}\right)=\operatorname{tr}[N^{\alpha,\beta}\left(\mathbf{k}\right)]$ 
for $J_2/J_1=0.25$ are displayed in Fig.~\ref{fig:observables}(b). Also in this case, 
we find a clear peak at the $\Gamma$ point. In order to understand whether this 
corresponds to diagonal order, we performed a finite-size scaling analysis 
of the ratio $N_0/V$, where $N_0$ is the largest eigenvalue of the diagonal two-body 
correlation function ($\langle n_i n_j \rangle -\langle n_i\rangle \langle n_j\rangle$ 
in the bosonic language, 
$\langle S^{z}_i S^{z}_j \rangle - \langle S^{z}_i\rangle \langle S^{z}_j\rangle$ in 
the spin language). Results for $N_0/V$, are shown in 
Figs.~\ref{fig:observables}(c)--\ref{fig:observables}(e) 
for $J_2/J_1=0.20,\, 0.25$ and $J_2/J_1=0.3$, respectively. They extrapolate to zero 
in the thermodynamic limit, which means that there is no charge order associated with 
the peak of $N\left(\mathbf{k}\right)$. Hence we conclude that the ground state
for $0.20\lesssim J_2/J_1\lesssim0.3$ is likely to be a spin liquid.

A comparison of the results of this study with those of 
Ref.~\cite{Varney2011}, reveals a very good agreement between the phase 
diagrams reported. However, there are important differences. The 
resulting momentum distribution in Fig.~\ref{fig:observables}(a) does 
not exhibit a Bose surface, but instead a clear peak at $\Gamma$. 
Since such trial states are known to be capable of reproducing Bose surfaces, 
one possibility is that the Bose surface seen in the exact calculations will 
fade away as the system size increases. However, since the absence of a 
Bose surface in our ansatz holds also for small system sizes, another 
possibility is that we have not found the trial state that is able to 
capture such a feature, and which may ultimately exhibit a lower energy 
than the one reported here. 

At this stage, the nature of the phase that we refer to as a ``spin 
liquid'' remains largely a mystery. The original exact diagonalization 
study \cite{Varney2011} found a state that was consistent with a 
gapless spin liquid. This scenario is still in the running, however, 
the cluster sizes available to such studies are too small to make an 
unambiguous conclusion. On the other hand, our results for state 
\eqref{eq:meanfield} are consistent with a gapped spin liquid, 
clearly in contrast to the projected spinon Fermi surface state 
suggested in Ref.~\cite{Varney2011}. In spite of the difficulties, 
in this work we have been able to safely rule out straightforward 
magnetically-ordered phases and spirals as competing ground states 
in the interesting region where the fractionalized spin-liquid state 
was found to win.

We note that it might be possible to reconcile the differences
observed by recalling the arguments in Ref.~\cite{Galitski2007}. 
There, it was suggested that, just like a Landau Fermi liquid can 
be viewed as a parent state subject to various instabilities 
(e.g., superconductivity, Stoner ferromagnetism, or density waves), 
the U$\left(1\right)$ projected Fermi liquid may serve as a parent 
state to a variety of Mott insulating states and gapped spin liquids.
Such scenarios have been discussed in the literature: a density wave 
instability~\cite{Altshuler1995} and a superconducting instability 
\cite{Galitski2007}, which may be consistent with the state we discuss 
here by virtue of its close relation with projected BCS states 
previously studied in Refs.~\cite{Lu2011,Clark2011,Gong2013}. 
In both cases, the nature of the possible instabilities hinges on 
the Friedel-like oscillations (or Kohn anomaly), which is produced 
by the fermionized bosons in the parent state. The existence of such 
oscillations can potentially be probed in numerical simulations of 
models with defects and would shed light on the true nature of the 
underlying state.

In summary, we have performed a variational study of the frustrated 
$XY$ model on the honeycomb lattice. We have been able to map out 
all ordered phases and describe them with variational states with an
unprecedented precision on lattices that for practical purposes 
represent thermodynamic limit (to which we believe there is no analog 
in the existing literature). We have also found that in the region 
$0.2\lessapprox J_1/J_2\lessapprox 0.3$, the ordered phases lose in 
energy to an exotic fractionalized partonic wave-function that is 
consistent with a gapped spin liquid. However, more work is 
needed to conclusively determine the true physical nature of the 
ground state of the system in this region.

This work was supported by the U.S. Office of Naval Research (J.C. and M.R.), the
National Science Foundation under Grant No.~OCI-0904597 (A.D.C. and M.R.),
NSF CAREER Award No. DMR- 0847224 (V.G.), JQI-PFC (A.D.C.), and PRIN 2010-11 (F.B.). 
We thank Alberto Parola for sharing his exact diagonalization results, and 
Steve White and Olexei Motrunich for discussions.

%

\newpage

\onecolumngrid


\begin{center}

{\large \bf Supplementary Materials:
\\Nature of the phases in the frustrated $XY$ model on the honeycomb lattice }\\

\vspace{0.6cm}

Juan Carrasquilla$^{1,2}$, Andrea Di Ciolo$^{3,2}$, Federico Becca$^{4,5}$, 
Victor Galitski$^3$, and Marcos Rigol$^1$ \\
\ \\
$^1${\it Department of Physics, The Pennsylvania State University, 
University Park, Pennsylvania 16802, USA}\\
$^2${\it Department of Physics, Georgetown University, Washington DC, 20057, USA}\\
$^3${\it Joint Quantum Institute and Department of Physics, 
University of Maryland, College Park, Maryland 20742, USA}\\
$^4${\it Democritos Simulation Center CNR-IOM Istituto Officina dei Materiali, Trieste, Italy }\\
$^5${\it International School for Advanced Studies (SISSA), Via Bonomea 265, 34136 Trieste, Italy }\\

\end{center}

\vspace{0.6cm}

\twocolumngrid

\subsection{Summary of the energies of different trial states on small clusters}

We have investigated different trial states based on the projected fermionic ansatz described 
in the main text with the aim of improving the results for the energy in the intermediate 
$J_2/J_1$ region. The results for the states with the lowest energies are listed in 
Tables~\ref{table:energies0.25} and~\ref{table:energies0.30} for $J_2/J_1=0.25$ and $0.3$,
respectively. 

The simplest state contains real NN hopping only, with 
$t^{\uparrow}_{ij}=-t^{\downarrow}_{ij}=t=1$; the mean-field Hamiltonian has gapless 
excitations with two Dirac points for each parton. The projected wave function has a 
relatively good accuracy, namely $\approx 2\%$ (for $J_2/J_1=0.25$) and $\approx 5\%$ 
(for $J_2/J_1=0.3$) on the $4 \times 4 \times 2$ cluster. We also added a staggered 
potential $\pm M$ on sublattice $\mathcal{A}$ ($+$ sign for parton $\uparrow$ and $-$  
sign for parton $\downarrow$), and $\mp M$ on sublattice $\mathcal{B}$, i.e., 
the sign is interchanged with respect to parton species. Moreover, we considered states 
with spin-orbit couplings $t^{s}_{ij}$ and $t^{a}_{ij}$, also allowing for ansatze 
that break PH symmetry. For particle-hole symmetric states, we considered states with real 
NN hopping $t^{\uparrow}_{ij}=-t^{\downarrow}_{ij}=t=1$, complex on-site, and real NNN 
spin-orbit hoppings, i.e., $m^{\uparrow \downarrow}_i$, and $t^{s}_{ij}=t^{s*}_{ij}$. 
As for states that break PH symmetry, we added (to the state just mentioned) non-zero 
NN and NNN $t^{a}_{ij}$ terms. 

The energy of the states that break PH symmetry presented in Tables~\ref{table:energies0.25} 
and~\ref{table:energies0.30}. Some of those states have slightly lower energy than the best 
PH symmetric states. The largest gain is of the order of $10^{-3}J_1$ for $J_2/J_1=0.3$. 
Despite this small energy gain, the physical picture discussed in the main text remains 
intact, in the sense that the single-particle gap remains finite and both $n_0/V$ and $N_0/V$ 
scale to values that are consistent with the results obtained for the states that preserved PH 
symmetry for most values of $J_2/J_1$. 

\begin{table}[ht]
 \caption{Energies of the different trial states and linear system sizes at $J_2/J_1=0.25$ 
and $l\times l \times 2$ clusters. H stands for hopping, ST for staggered potential as 
described in the text, and SO for ``spin-orbit'' coupling.}

 \centering
 \begin{tabular}{l l l}
  \hline\hline
  $l$ & State & Energy  \\ [0.5ex]
  \hline
  4  &  Exact diagonalization                                        &   $-0.302285$    \\
  4  &  Real NN H                                                    &   $-0.29595(1)$  \\
  4  &  Real NN H + ST                                               &   $-0.29650(1)$  \\
  4  &  Real NN + SO                                                 &   $-0.29654(1)$  \\
  4  &  4-site unit cell + ST                                        &   $-0.29650(1)$  \\
  4  &  Real NN + ST + breaking $\mathcal{J}_z$                      &   $-0.29649(1)$  \\
  4  &  Real NN + SO breaking PH symmetry                            &   $-0.29653(1)$  \\
  \hline
  6  &  Real NN H                                                    &   $-0.294751(8)$ \\
  6  &  Real NN + ST                                                 &   $-0.294972(7)$ \\
  6  &  Real NN + SO                                                 &   $-0.295294(5)$ \\
  6  &  18-site unit cell + ST (see pattern in Fig.1)                &   $-0.294984(6)$ \\
  6  &  18-site unit cell + ST unconstrained                         &   $-0.29498(1)$  \\       
  6  &  Real NN + SO breaking PH symmetry                            &   $-0.295789(3)$ \\
  6  &  AF state + breaking $\mathcal{J}_z$                          &   $-0.281708(4)$ \\              
  6  &  AF state + on-site potential + breaking $\mathcal{J}_z$      &   $-0.281686(7)$ \\
  6  &  CL state +  breaking $\mathcal{J}_z$                         &   $-0.276165(5)$ \\
  6  &  CL state + on-site potential + breaking $\mathcal{J}_z$      &   $-0.276135(8)$ \\
  \hline
 \end{tabular}

 \label{table:energies0.25} 

\end{table}

In order to assess the breaking of spatial symmetries, we considered various mean-field 
Hamiltonians with enlarged unit cells. On a small  $4\times4\times2$-site cluster, we considered 
systems with independent hopping and staggered on-site parameters (that respect PH symmetry) 
in a $4$-site unit cell. Note that such states could potentially attain charge-density wave 
(CDW) states by virtue of the independence of all on-site parameters in the wave function. 
However, no CDW state was found in our numerical optimizations: after the numerical optimization 
a translationally invariant state was obtained. All states that we have mentioned so far were 
supplemented with a rotationally symmetric Jastrow factor $J_z$ (allowing non-rotational 
invariant Jastrow factors does not lead to a sizable energy gain). 

\begin{table}[ht]
 \caption{Energies of the different trial states and linear system sizes at $J_2/J_1=0.3$ 
and $l\times l \times 2$ clusters. H stands for hopping, ST for staggered potential as described 
in the text, and SO for ``spin-orbit'' coupling.} 

 \centering  
 \begin{tabular}{l l l} 
  \hline\hline          
  $l$ & State & Energy  \\ [0.5ex] 
  \hline                  
  4  &  Exact diagonalization                                        &   $-0.295275$   \\ 
  4  &  Real NN H                                                    &   $-0.28043(2)$ \\ 
  4  &  Real NN H + ST                                               &   $-0.28221(1)$ \\
  4  &  Real NN + SO                                                 &   $-0.28222(1)$ \\
  4  &  4-site unit cell + ST                                        &   $-0.28221(1)$ \\
  4  &  Real NN + ST + breaking $\mathcal{J}_z$                      &   $-0.28211(1)$ \\
  \hline 
  6  &  Real NN H                                                    &   $-0.279596(2)$ \\
  6  &  Real NN + ST                                                 &   $-0.280840(8)$ \\
  6  &  Real NN + SO                                                 &   $-0.281169(7)$ \\
  6  &  18-site unit cell + ST (see pattern in Fig.1)                &   $-0.280849(4)$ \\
  6  &  18-site unit cell + ST unconstrained                         &   $-0.280989(4)$ \\ 
  6  &  AF state + breaking $\mathcal{J}_z$                          &   $-0.260883(8)$ \\ 
  6  &  AF state + breaking $\mathcal{J}_z$ + on-site potential      &   $-0.260889(1)$ \\
  6  &  CL state + breaking $\mathcal{J}_z$                          &   $-0.278283(4)$ \\
  6  &  CL state + breaking $\mathcal{J}_z$ + on-site potential      &   $-0.278275(4)$ \\
  \hline
 \end{tabular}

 \label{table:energies0.30} 

\end{table}

\begin{figure}[!t]
\includegraphics[width=0.43\textwidth]{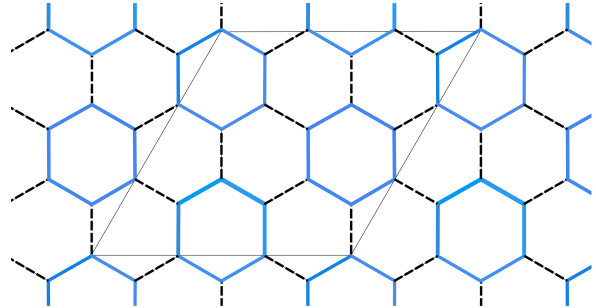}
\caption{(Color online) Assumed pattern of the hopping amplitudes for the state with a 
$18$-site unit cell and two types of hopping. The unit cell is depicted as a parallelogram 
using thin solid black lines.}
\label{fig:18siteunitcell}
\end{figure}

For the $6\times6\times2$ cluster we also explored the breaking of spatial symmetry on
states constructed on a larger $18$-site unit cell, inspired by the plaquette-like phases 
found in Ref.~\cite{Ganesh2013,Zhu2013} for the Heisenberg model. The first state we 
considered was constructed by defining two different real NN hopping amplitudes assigned 
to different bonds as depicted in Fig.~\ref{fig:18siteunitcell}, e.g., $t_1$ for continuous 
blue bonds and $t_2$ for dashed black bonds. Furthermore, we allowed all $18$ on-site potentials 
in the unit cell to be optimized independently in order to search for possible CDW states. Upon 
optimization, we recovered a fully symmetric state even if we allowed for the presence of a 
staggered potential with interchanged signs and the same magnitude, as explained before. 
If we release the constrains and optimize all $27$ bonds in the unit cell, we find a slight 
energy gain with respect to the homogeneous case with real NN hoppings, though the energy 
is still higher than the homogeneous states with spin-orbit couplings. 

Finally, let us discuss the trial states used to describe classically ordered 
states. Also in this case, we considered terms that break translational invariance, i.e., 
a site-dependent one-body Jastrow factor on top of $\mathcal{J}_z$:
\begin{equation}
|\Psi_{\mathbf{Q}}\rangle=\mathcal{J}^{loc}_z\mathcal{J}_z \prod_{i} \left(|\downarrow\rangle_{i} + 
e^{\imath\mathbf{Q}.\mathbf{R}_i+\imath \eta_{\mathbf{R}_i}}|\uparrow\rangle_{i} \right),
\label{eq:spirals_local}
\end{equation}
where $\mathcal{J}^{loc}_z=\exp{\left (\sum_i v_i S^{z}_i \right)}$ and $v_i$ are optimized 
independently for each site $i$ in the clusters. In particular, we considered the energy 
gain in both antiferromagnet (AF) and collinear (CL) states upon introduction of the on-site 
term for $J_2/J_1=0.25$ and $0.3$. In spite of the on-site term, no gain in energy and no 
evidence of breaking of translational invariance or CDW state was found.  

\subsection{Accuracy of the best trial states on small clusters}

\begin{figure}[!t]
\includegraphics[width=0.23\textwidth,angle=-90]{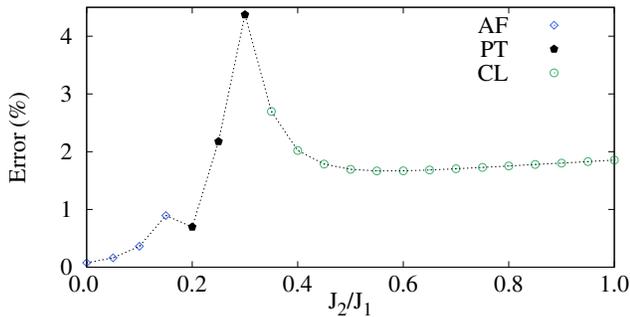}
\caption{(Color online) The percentile error with respect to results from exact 
diagonalization in the energies of the antiferromagnetic (AF), partonic (PT), and collinear (CL) 
wave functions.}
\label{fig:error}
\end{figure}

In Fig.~\ref{fig:error} we present the error in the energy of our best ansatze, i.e.,
antiferromagnetic, collinear and partonic states, with respect to results from exact 
diagonalization on the $4 \times 4 \times 2$ cluster.The discrepancy in the energy is 
always less than $3\%$, except for $J_2/J_1=0.3$ where a slightly lower accuracy is 
observed $\sim 4\%$. Remarkably, the error in the ordered states is typically less than 
$2\%$. This gives us confidence that such states provide an excellent approximation to 
the ground states of the model under consideration in the relevant regimes. More 
importantly, we can safely exclude incommensurate spirals in the intermediate 
regime where their energies lose against the partonic state. This is important
because excluding incommensurate spirals is difficult in exact diagonalization 
studies due to the limitations in the cluster sizes that can be handled.
 
\subsection{Energies of the trial states and extrapolations to the thermodynamic limit}

In Table ~\ref{table:energies} we present the energies of the best trial states for some 
$l\times l \times 2$ cluster and values of $J_2/J_1$, as well as some of the extrapolations 
to the thermodynamic limit presented in Fig.~3 of the main text. The first 2 values of 
$J_2/J_1$ (i.e., $0.10$, $0.15$) correspond to the classical AF state supplemented with 
$\mathcal{J}_z$. The next 3 values (i.e., $0.20$, $0.25$, and $0.30$) correspond to the best
projected fermionic state. Finally, the last 3 values (i.e., $0.35$, $0.40$, and $0.45$) correspond 
to the CL state also supplemented with $\mathcal{J}_z$.  

\onecolumngrid

\begin{table*}[!h]
\squeezetable 
 \caption{Energies of the best trial states for some cluster sizes $l\times l \times 2$  and values 
of $J_2/J_1$.  The last row shows some of the extrapolated energies to the thermodynamic limit 
presented in Fig.~3 of the main text. The errors in the finite-sized clusters are statistical errors 
from the variational Monte Carlo procedure, while the errors of the data in the thermodynamic limit 
come from  the fitting procedure used in the extrapolations.}
 \centering  
 \begin{tabular}{c c c c c c c c c} 
  \hline\hline          
  $l$    & $J_2/J_1=0.10$ & $0.15$          & $0.20$         & $0.25$         & $0.30$         & $0.35$         & $0.40$         & $0.45$ \\
  \hline                  
  4      & $-0.364116(3)$ & $-0.334322(4)$  & $-0.312664(7)$ & $-0.29654(1) $ & $-0.28222(1) $ & $-0.286302(3)$ & $-0.295509(3)$ & $-0.306309(3)$ \\
  6      & $-0.362540(2)$ & $-0.333060(2)$  & $-0.311063(4)$ & $-0.295294(5)$ & $-0.281169(7)$ & $-0.285055(3)$ & $-0.294064(3)$ & $-0.304612(3)$ \\
  12     & $-0.361960(2)$ & $-0.332590(3)$  & $-0.311141(3)$ & $-0.295464(3)$ & $-0.281552(4)$ & $-0.284592(2)$ & $-0.293546(2)$ & $-0.304026(2)$ \\
  18     & $-0.361903(1)$ & $-0.332541(2)$  & $-0.311129(3)$ & $-0.295461(4)$ & $-0.28156(2) $ & $-0.284543(2)$ & $-0.293495(2)$ & $-0.303968(2)$ \\
$\infty$ & $-0.36188(1) $ & $-0.33253(2) $  & $-0.31107(3) $ & $-0.29541(4) $ & $-0.28154(3) $ & $-0.284526(2)$ & $-0.293472(2)$ & $-0.30393(1) $ \\
  \hline 
 \end{tabular}

 \label{table:energies} 

\end{table*}


\end{document}